# Valley-protected backscattering suppression in silicon photonic graphene

Xiao-Dong Chen and Jian-Wen Dong*

*School of Physics & State Key Laboratory of Optoelectronic Materials and Technologies, Sun Yat-Sen University, Guangzhou 510275, China.*

*Corresponding author: <dongjwen@mail.sysu.edu.cn>

**Abstract**

**In this paper, we study valley degree of freedom in all dielectric silicon photonic graphene. Photonic band gap opening physics under inversion symmetry breaking is revisited by the viewpoint of nonzero valley Chern number. Bulk valley modes with opposite orbital angular momentum are unveiled by inspecting time-varying electric fields. Topological transition is well illustrated through photonic Dirac Hamiltonian. Valley dependent edge states and the associated valley-protected backscattering suppression around Z-shape bend waveguide have been demonstrated.**

Graphene, a two-dimensional honeycomb lattice of carbon atoms, supports Dirac cones in low energy electronic band structure [1]. This conical band dispersion leads to many profound phenomena such as Klein tunneling [2], Zitterbewegung [3], and integer quantum Hall effect [4]. Recently, photonic graphene, the optical counterpart of graphene in electronics, has been utilized to study much predominant photonic behaviors [5-8]. A representative example is topological photonic crystals that have a surprising ability for molding the flow of light [9-23]. Several fundamental nontrivial behaviors such as spin-dependent nonzero Berry phases, spin-directional locking behaviors and synthetic gauge fields have been illustrated [10-14]. Some revolutionary applications such as one-way waveguide, robust optical delay line and robust transport channel have been observed [12-19]. However, either breaking time-reversal symmetry or complicated metamaterial structures are required in previous literatures, limiting the realistic implementation of topological photonic crystals. Until very recently, a new kind of topological photonic crystal relies on the $C_6$ crystal symmetry is proposed in an all dielectric platform [23].

In this paper, we study the valley degree of freedom in a time-reversal invariant photonic graphene, constructed by all dielectric silicon honeycomb crystal. Photonic band gap opening is revisited by the viewpoint of nonzero valley Chern number. The orbital angular momentum of photonic modes in the bulk crystal is well exploited. Topological transition is well demonstrated in the silicon photonic graphene, derived from the photonic Dirac Hamiltonian. Valley dependent edge states and the associated broadband high transmission around Z-shape bend waveguide have been demonstrated.

Consider a two-dimensional photonic graphene, which is a composite of two interlaced triangular lattices with same lattice constant (Fig. 1a). The unit-cell consists of two silicon rods embedded in the air background (A and B sub-lattice in dash-rhomboid or dash-hexagon). Each rod has a diameter of $d = 0.44a$ where $a$ is the lattice constant. The relative permittivity is $\varepsilon = 11.7$ corresponding to that of silicon behaves at the wavelength of 1550nm. Figure 1b shows the lowest bulk bands of transverse magnetic modes with nonzero z-component of electric field (grey curves). Conical dispersions with Dirac frequency of 0.2574 $c/a$ appear at two inequivalent but time-reversal valleys at Brillouin zone corner (K' and K). Note that the group symmetry of honeycomb lattice yields the Dirac dispersions without fine tuning of physical parameters. The conical dispersions at the two inequivalent valleys are protected by the combination of the spatial inversion symmetry and time-reversal symmetry [24]. It will be gapped out immediately when one of symmetry is broken. In order to illustrate it, we change the diameter of two rods ($d_A$ and $d_B$) while keep the average diameter $(d_A + d_B)/2$ unchanged. The parameter of $\delta d = (d_A - d_B)/2$ is defined to describe the amplitude of inversion symmetry breaking. Black curves in Fig. 1b show the two lowest bulk bands of photonic graphene with $\delta d = 0.06a$. It is clear to see that Dirac cones are split out and a complete band gap appears from the frequency of 0.244 to 0.272 $c/a$. From the group theory point of view, when the inversion symmetry is broken, the group symmetry of K' (or K) point is reduced from $C_{3v}$ to $C_3$. The degenerate irreducible representation $E$ will transform into two non-degenerate irreducible representations $^1E$ and $^2E$ [25]. This is verified in Fig. 2 that the $E_z$-fields of valley modes in band 2 (band 1) concentrate in the B (A) rod which has smaller (larger) radius. Hence valley modes at band 2 should have higher frequency than those at band 1, and consequently they will be separated in the frequency level.

But this is not the end of the story. We emphasize an interesting issue such that each valley mode of Dirac band has an intrinsic orbital angular momentum (OAM) of time-varying $E_z$ fields. Take the valley mode of band 1 at K' point for example, as shown in Fig. 2c. As time evolves, the $E_z$ fields rotate counterclockwise (see the animation in Supplementary Information), which is consistent with the Poynting vectors circle counterclockwise around the origin (green arrows in Fig. 2c). We claim that such valley mode has a left circular polarized (LCP) OAM. The other three valley modes can be also identified according to the corresponding information of OAM. In order to have a deep understanding on the role of OAM in the photonic valley modes, we construct a minimal band model of Dirac dispersions. Through the **k·p** approximation, we can resort to the Dirac Hamiltonian:

$$\hat{H} = v_D(\hat{\sigma}_x \hat{\tau}_z \delta k_x + \hat{\sigma}_y \delta k_y) + \lambda^P_{\varepsilon_z} \hat{\sigma}_z \quad (1)$$

where $\delta \bar{k}$ is measured from the valley center K' or K. $\hat{\sigma}_i$ and $\hat{\tau}_i$ are the Pauli matrices acting on sub-lattice and valley spaces, respectively. In particular, $2\lambda^P_{\varepsilon_z}$ determines the frequency gap at valley centers with $\lambda^P_{\varepsilon_z} \propto [\varepsilon_z(B) - \varepsilon_z(A)]$ where $\varepsilon_z(*)$ denotes the integration of $\varepsilon_z$ at A or B rod [see detailed derivation in Supplementary Information]. As a result, $\lambda^P_{\varepsilon_z}$ will be nonzero and band gap will open due to the inversion asymmetric $\varepsilon_z$. Moreover, the Dirac Hamiltonian of Equation (1) implies the Berry curvature has a distribution centered at two valleys, $\Omega(\delta \bar{k}) = \tau_z \cdot 3\lambda^P_{\varepsilon_z} / [(2\lambda^P_{\varepsilon_z})^2 + 3(\delta \bar{k})^2]^{3/2}$, and each valley carries a nonzero topological charge with opposite signs $C_{\tau_z} = \tau_z \,\mathrm{sgn}(\lambda^P_{\varepsilon_z})/2$ [26]. As a result, it does acquire the nonzero topological invariant -- valley Chern number $C_v = (C_K - C_{K'}) \neq 0$, although the total Chern number $C = (C_K + C_{K'})$ is zero for the time-reversal invariant silicon photonic graphene.

Another interesting issue is the topological transition in the evolution of the inversion asymmetry parameter $\delta d$. The topological charge distributions for the silicon photonic graphene with $\delta d < 0a$ and $\delta d > 0a$ are totally different. Left panel of Fig. 1d illustrates that the topological charge is +1/2 at K valley and -1/2 at K' valley, when $\delta d < 0a$, resulting that the silicon photonic graphene with $\delta d < 0a$ has a nonzero positive valley Chern number of $C_v = 1$. On the other hand, the signs of topological charge for $\delta d > 0a$ are flipped (right panel of Fig. 1d), and thus the silicon photonic graphene has a nonzero negative valley Chern number of $C_v = -1$.

Figure 1c plots the evolution of the valley modes as a function of $\delta d$. With the increase of $\delta d$, the radius of A rod becomes larger but that of B rod becomes smaller (see in upper insets of Fig. 1c). As a result, those modes with electric fields concentrating at A (B) rod will has lower (higher) frequency. So the frequency spectra for LCP K' valley mode and RCP K valley mode drop monotonously, while those for the other two valley modes boost up as the increasing of $\delta d$. More importantly, exchange between modes at same valley happens at $\delta d = 0a$, and causes the topological transition. It gives potential to the valley dependent edge states by interfacing two different silicon photonic graphene together.

To see the valley dependent edge states, we focus on the edges with zigzag morphology. Consider the zigzag edge (cyan line in Fig. 3b) constructing by silicon photonic graphene with $\delta d = 0.06a$ at the bottom and another with $\delta d = -0.06a$ at the top. As have been shown above, the silicon photonic graphene with $\delta d = 0.06a$ ($\delta d = -0.06a$) has a topological charge of +1/2 (-1/2) at K' valley and -1/2 (+1/2) at K valley. Then the local topological charge differences across the edge at K' and K valleys are +1 and -1, respectively. According to the bulk-edge correspondence [27], there will be one edge state with positive velocity at K' valley and one with negative velocity at K valley. These valley dependent edge dispersions are confirmed by the simulated results shown in Fig. 3a. In addition, all edge states can be classified into odd or even parities as the edge structure has mirror symmetry. Two representative $E_z$ fields of the edge states, corresponding to the black circles in the cyan edge dispersions, are shown in the right insets. It tells us that the lower edge states are odd modes while the upper edge states are even modes. Moreover, the edge dispersions can be controlled by interchanging the relative positions of two silicon photonic graphene (Fig. 3d). This edge consists of silicon photonic graphene with $\delta d = -0.06a$ at the bottom while that with $\delta d = 0.06a$ at the top. Hence, the topological charge differences at valleys are flipped. As a result, the edge states at K' valley flow to negative direction while those at K valley flow to positive direction (Fig. 3c). Once again, all edge states can be classified into odd or even parities (right insets of Fig. 3c). However, it should be noted that the edge states do not have the same origin as those in photonic QHE by breaking time-reversal symmetry [15] or photonic QSHE by introducing photonic spin-orbital coupling [12]. The edge dispersions in Fig. 3 do not connect the valence and conduction bulk bands and they are gapped.

Although the valley dependent edge states of silicon photonic graphene are topologically trivial as the global Chern number is zero, they can be employed for constructing high transmission bend waveguide for the reason of the inter-valley scattering suppression protected by the local topological distinction at the two inequivalent valleys [28]. Figure 4a shows the schematic diagram on the Z-shape bend waveguide constructed by the zigzag edge in Fig. 3d. A mirror symmetric $E_z$ line sources is set from the top left. The transmittance of the Z-shape bend waveguide and the comparative straight channel are measured, and marked with the red and black transmission spectra in Figure 4b, respectively. The frequency region of the edge dispersions with even parities are shaded in light-pink rectangle. As we can see, the transmission of the Z-shape bend is lower than that of the straight channel from 0.237 to 0.244 $c/a$. This can be well understood by noticing that the bulk modes of photonic graphene exist in this frequency region. When the propagating electromagnetic waves meet the bend corners, the waves will be scattered into the bulk crystal, see e.g. the electric density distribution at 0.24 $c/a$ in Fig. 4c. This causes the loss of energy and leads to low transmission. In contrast, when the frequency of the edge states locates at the complete band gap of silicon photonic graphene, the scattering into bulk modes is forbidden. In addition, the backscattering between the forward and backward edge states is also blocked due to the suppression of inter-valley scattering. Therefore, broadband high transmission can be achieved. As shown in Figs. 4d-4f, the electromagnetic waves can pass through the Z-sharp bend waveguide without backscattering. Finally, the transmission through Z-sharp bend is almost the same as that of straight channel and a broadband high transmission is achieved in Fig. 4b.

In conclusion, we study the bulk states and topology inside the time-reversal silicon photonic graphene. The band gap opening physics is revisited by the viewpoint of valley Chern number. The orbital angular momentum of bulk valley modes are well unveiled by the eigen-fields analysis. Topological transition is verified by the changing of the local topological charge distributions and total valley Chern number. Valley dependent edge states are found and the associated high transmission around Z-shape bend is demonstrated. Our work may open up a new route towards the discovery of fundamentally novel states of light in silicon optics.

This work is supported by the Natural Science Foundation of China (11274396, 11522437), the Guangdong Natural Science Funds for Distinguished Young Scholar (S2013050015694), the Guangdong special support program, and the SYSU visiting scholarship for the UC Berkeley period.

**Figures and Figure Captions**

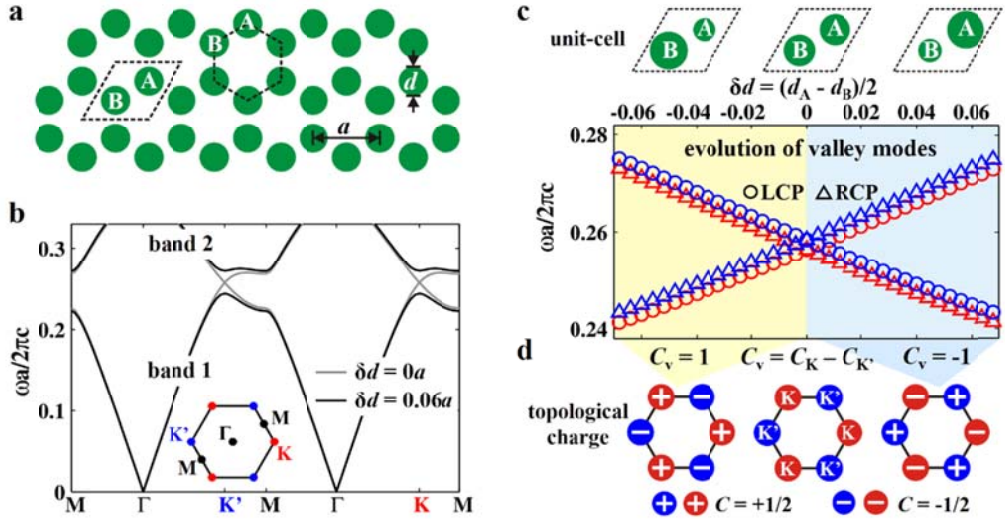

**Figure 1. Inversion symmetry breaking silicon photonic graphene with nonzero valley Chern number. a**, Schematic of silicon photonic graphene. The unit-cell consists of two silicon rods embedded in the air background (A and B sub-lattice in dash-rhomboid or dash-hexagon). Each rod has a diameter of $d_A = d_B = 0.44a$ where $a$ is the lattice constant and the relative permittivity of $\varepsilon = 11.7$. **b**, The lowest bulk bands of transverse magnetic modes with nonzero z-component of electric fields for silicon photonic graphene with $\delta d = 0a$ (grey curves) and with $\delta d = 0.06a$ (black curves). Here, the parameter of $\delta d = (d_A - d_B)/2$ is defined to describe the amplitude of inversion symmetry breaking. Dirac cones appear at two inequivalent but time-reversal valleys and they are gapped out when the inversion symmetry is broken. The first Brillouin zone with high symmetry $k$-points is also shown in inset. **c**, Evolution of valley modes as a function of $\delta d$. Frequency spectra for LCP K' valley mode and RCP K valley mode drop monotonously, while those for the other two valley modes boost up as the increasing of $\delta d$. Mode exchange happens at $\delta d = 0a$ and results in the topological transition. **d**, Schematics of topological charge distributions for silicon photonic graphene with $\delta d < 0a$ and $\delta d > 0a$. Nonzero valley Chern numbers, i.e., $C_v = (C_K - C_{K'})$ which distinguish the topology, are found and they are different for two distinct silicon photonic graphene. It indicates the topological transition and gives potential to the valley dependent edge states by interfacing two different silicon photonic graphene together.

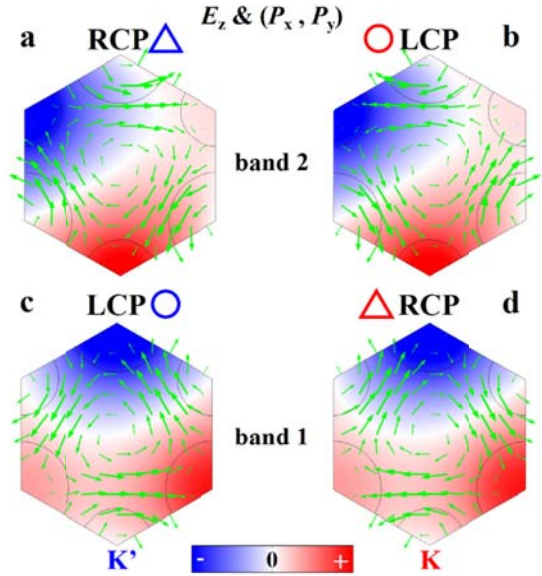

**Figure 2. Valley modes with circular polarized orbital angular momentum.** The z-component of electric fields ($E_z$) and in-plane Poynting vectors ($P_x$, $P_y$) are shown. **a-b**, Eigen-fields of valley modes at band 2 at (**a**) K' valley and (**b**) K valley. The $E_z$ fields between them are the same due to the time-reversal symmetry. **c-d**, Eigen-fields at valley modes of band 1 at (**c**) K' valley and (**d**) K valley. The $E_z$ fields of valley modes at band 2 (band 1) concentrate in the B (A) rod which has smaller (larger) radius. Therefore valley modes at band 2 should have higher frequency than those at band 1, and they will be separated in the frequency level. In addition, take valley mode of band 1 at K' valley in (**c**) as an example, the $E_z$ fields rotate counterclockwise as time evolves (see the corresponding animation in Supplementary Information). The Poynting vectors also circle counterclockwise around the origin. So this valley mode has left-handed circular polarized (LCP) orbital angular momentum (OAM) of $E_z$ fields. By the same procedure, other three valley modes are also classified according to the orbital information.

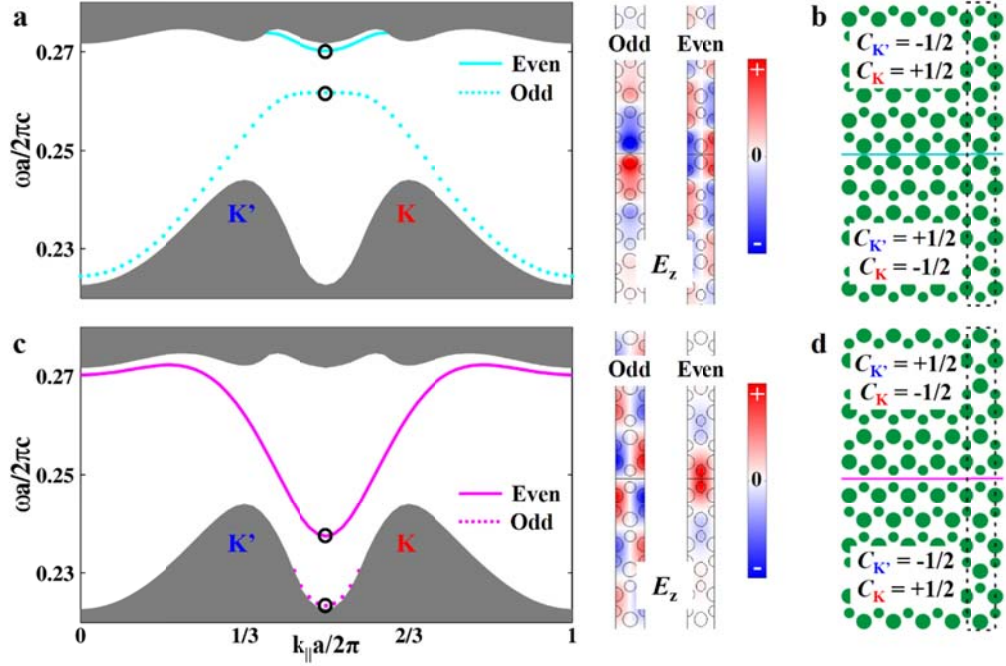

**Figure 3. Valley dependent edge states**. **a**, Valley dependent edge states, i.e., one edge state with positive velocity at K'valley and one with negative velocity at K valley, are found. **b**, The corresponding zigzag edge is constructed by silicon photonic graphene with δ$d$ = 0.06$a$ on the bottom and silicon photonic graphene with δ$d$ = -0.06$a$ on the top. In addition, all edge states can be classified into odd or even parities as the edge structure has mirror symmetry. Representative $E_z$ fields of odd and even edge states, corresponding to black circles in the cyan edge dispersions, are shown in the right insets. **c**, Valley dependent edge states with flipping group velocities in each valley are realized in **d**, interchanging the relative positions of two silicon photonic graphene.

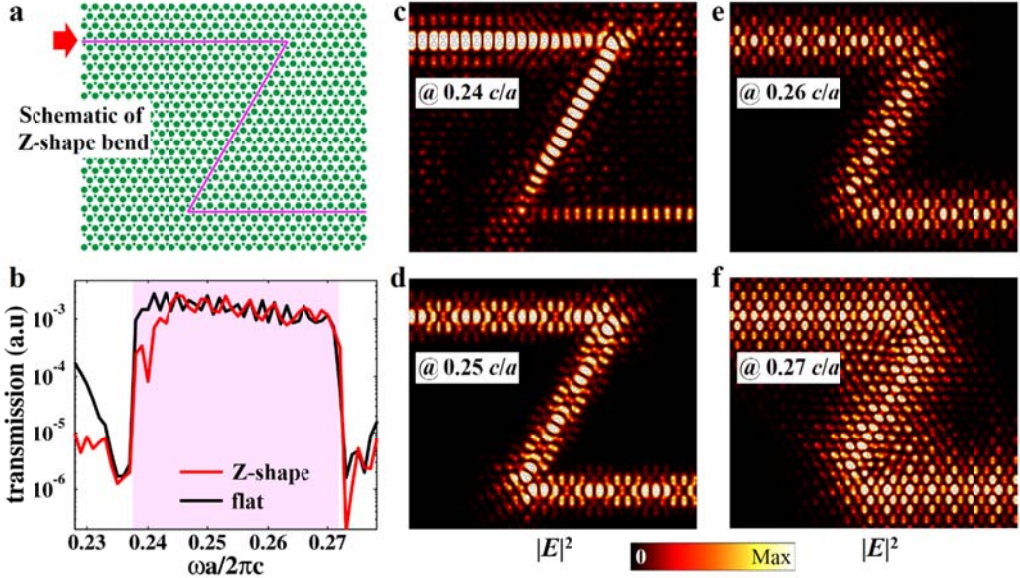

**Figure 4. High transmission around Z-shape bend**. **a**, Schematic of the Z-shape bend. Mirror symmetric $E_z$ line sources are incident from the top-left end, and transmission at the bottom-right end is measured. The transmission of flat channel without Z-shape bend is also measured for comparison. **b**, Transmissions of Z-shape bend and flat channel are shown in red and black curves, respectively. From 0.237 to 0.244 *c/a*, transmission of Z-shape bend is lower than that of flat channel as the edge states are scattered into the bulk silicon photonic graphene. **c**, The electric density distribution at 0.24 *c/a*, showing that the excited electromagnetic waves are scattered when they meet the bend corners. **d-f**, The electric density distributions at (**d**) 0.25 *c/a*, (**e**) 0.26 *c/a* and (**f**) 0.27 *c/a*, illustrate that the excited electromagnetic waves can pass through the sharp bend corners without backscattering as Z-shape bend does not induce inter-valley scattering. Hence the transmission through Z-shape bend is the same as that through flat channel and a broadband high transmission is achieved from 0.244 to 0.272 *c/a*, which can be seen in (**b**) with two transmission curves are stick together.